\documentclass[pre,showpacs]{revtex4-1}
\usepackage[utf8x]{inputenc}
\usepackage{amssymb,amsmath}
\usepackage[margin=1in]{geometry}
\usepackage{setspace}
\usepackage{natbib}
\usepackage{graphicx}
\usepackage{bm}

	 \renewcommand{\vec}[1]{\bm{#1}}

\begin{document}

%============================================================
\title{Fluctuation Theorem for Quasi-Integrable Systems}
%============================================================

\author{Tomer Goldfriend}
\email{tomergf@gmail.com}
\affiliation{Laboratoire  de  Physique  Statistique, D\'epartement de physique de l’ENS, \'Ecole  Normale  Sup\'erieure,
PSL  Research  University;  Universit\'e  Paris  Diderot, Sorbonne  Paris-Cit\'e;  Sorbonne
Universit\'es,  UPMC  Univ.  Paris  06,  CNRS;  24  rue  Lhomond,  75005  Paris,  France}

\author{Jorge Kurchan} 
\affiliation{Laboratoire  de  Physique  Statistique, D\'epartement de physique de l’ENS, \'Ecole  Normale  Sup\'erieure,
PSL  Research  University;  Universit\'e  Paris  Diderot, Sorbonne  Paris-Cit\'e;  Sorbonne
Universit\'es,  UPMC  Univ.  Paris  06,  CNRS;  24  rue  Lhomond,  75005  Paris,  France}

%============================================================
\date{\today}

\begin{abstract}
A Fluctuation Theorem (FT), both Classical and Quantum,  describes the large-deviations in the approach to equilibrium of an isolated quasi-integrable system.
Two characteristics make it unusual: {\em i)}  it concerns the internal dynamics of an isolated system without external drive, and {\em ii)} unlike the usual FT,
the system size, or the time,  need not be small for the relation to be relevant, provided the system is close to integrability.
 As an example, in the Fermi-Pasta-Ulam  chain, the relation gives information on  the ratio of probability of death to resurrection of solitons.
For a coarse-grained system the FT describes how the system `skis' down the (minus) entropy landscape: always descending but generically not along a gradient line. 
\end{abstract}

\maketitle

{\it Introduction.} --- Quasi-integrable systems--- those whose Hamiltonian slightly differs from an integrable one--- are widespread in Nature, ranging from planetary systems, to weakly nonlinear waves,  to some  quantum chains.  When the systems are macroscopic they often  involve coherent stable structures like solitons~\cite{Zakharov&Filonenko1967}.
A small breaking of integrability  leads to slow dynamics toward equilibrium~\cite{Benettin_etal2013}. The fluctuations and the nature of irreversibility along this slow route is the subject of this letter. The discussion is very similar for Classical and Quantum systems.

The trajectories of a classical integrable system with $N$ degrees of freedom correspond to a laminar flow along $N$-dimensional torus, defined by $N$ integrals of motion $\{J_r(\vec{x})\}^N_{r=1}$.
According to the Kolmogorov-Arnold-Moser (KAM) theorem, as the integrability-breaking interactions are switched on, some tori remain unbroken, or change exponentially in time according to Nekhoroshev theorem. However,  these intermediate regimes are expected to exist for  a range of  coupling parameters that is vanishingly small -- and is often irrelevant -- for macroscopic systems, which
  thus perform unimpeded diffusion towards equilibrium.  A description of weakly conserved quantities is given for, e.g., energy cascade in weakly coupled wave modes (weak turbulence)~\cite{Zakharov&Filonenko1967,During_etal2006}, or slow relaxation in a gas with long-range interactions~\cite{Chavanis2007}. 

A generic and somewhat surprising feature is that the Lyapunov time, measuring the separation of nearby trajectories, is in such systems often much shorter than the characteristic diffusion time of the quasi-constants of motion. Such is the case of  the Solar System~\cite{Laskar2008} (Lyapunov time of 5 Myrs $\ll$ 5 Gyrs for the stability time) and the Fermi-Pasta-Ulam-(Tsingou) (FPU) nonlinear chain~\cite{Benettin_etal2013,Benettin_etal2018,Goldfriend&Kurchan2018} (Lyapunov time of $10^6$ compared to thermalization time of $10^{10}$ for a chain of size N=1024). 
A simple way to rationalize  this is to consider the solvable problem of an integrable system perturbed by weak stochastic noise~\cite{Lam&Kurchan2014}:
 chaos develops first tangent to the action (invariant) variables, allowing the system to have ergodic motion primarily within a torus, and diffusion at a longer scale;
  there is furthermore no  KAM regime for any amplitude of the (white) noise.

We consider here systems with any number of quasi-conserved quantities, $J_1,...,J_p$, corresponding to any model with a finite number of constants of motion which is weakly perturbed. In addition, the systems may have strictly conserved quantities, such as the total energy or linear momentum, imposed by the symmetries of the complete perturbed Hamiltonian. We indicate these by $J^c_1,...,  J^c_n$ and assume that they are independent -- no function of them vanishes identically -- and that no function of the nonconserved  $J_1,...,J_p$ is strictly conserved \footnote{We may thus add a conserved $ J^c_a$  to any nonconserved $J_b$, but not a 
nonconserved $J_a$  to a conserved $ J^c_b$.}. We shall only distinguish these two sets when necessary, otherwise we drop the superscript.

The statistics of a system with many conserved quantities  may in some cases be described by a Generalized Gibbs Ensemble (GGE), 
\begin{equation}
\rho_{\rm GGE}(x)=e^{-Q(x)}/Z  \;\;\; \;\;\;  ; \;\;\; \;\;\; Z=\int dx e^{-Q(x)},
\label{GGE}
\end{equation}
where we define
\begin{equation}
Q(x)= \sum\beta_r J_r(\vec{x}) \label{Q}
\end{equation}
and $\{\beta_r\}$ are Lagrange multipliers. In the case when the $J_r$ are not true constants, the $\beta_r$ would be slowly time-dependent~\cite{Essler_etal2014,Stark&Koller2013}.
 The GGE construction  can be understood   as a form  of maximal entropy principle~\cite{Jaynes1957}, or, as a consequence of equivalence of ensembles (to be discussed below) in analogy to the Gibbs measure; see Ref.~\cite{Yuzbashyan2016}. Recently, the GGE distribution has  proved useful in describing quantum integrable systems~\cite{Vidmar&Rigol2016} (and references therein), and analogous classical ones~\cite{Cugliandolo2018}, where a direct access to the $\beta_r$ is granted by the response and correlation function~\cite{Foini_etal2017}.

Fluctuation theorems refer to a group of relations concerning a system which evolves under non-equilibrium conditions~\cite{Evans_etal1993,Evans&Searles1994,Gallavotti&Cohen1995,Kurchan1995,Lebowitz&Spohn1999,Maes1999,Evans&Searles2002,Crooks1999,Jarzynski1997}. 
Mainly, one considers a system driven away from an equilibrium state by an external noise or non-conservative forces, and studies the heat/entropy exchange or work extraction. Here, we study a {\it generalized exchange fluctuation theorem} (GXFT) for quasi-integrable systems. We consider a system which starts at a GGE state and study the fluctuations in the quasi-conserved quantities, $\Delta J_r$, as induced by the weak breaking of conservation. In particular, we define the quantity $u\equiv \sum\beta_r \Delta J_r(\vec{x})  $, and prove that it obeys the following relation
\begin{equation}
\ln \frac{P(u)}{P(-u)}=u.
\label{eq:GXFT}
\end{equation}

FT for this quantity in strictly (quantum) integrable systems, where work is performed by an external agent, have been considered recently in Refs.~\cite{Hickey&Genway2014,MurPetit_etal2018}. Here the situation is different, we are interested
in the endogenous entropy production due to integrability breaking.

\begin{figure}
\centerline{\resizebox{0.45\textwidth}{!}{\includegraphics{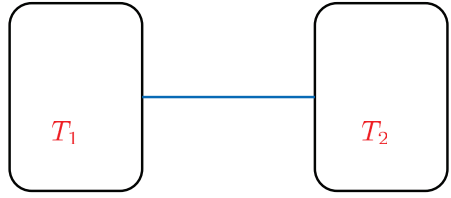}}}
%\hspace{2cm}}
\caption[]{The simplest GGE set-up, coinciding with the fluctuation theorem of Jarzynski and W\'ojcik~\cite{Jarzynski&Wojcik2004}. Two isolated equilibrated systems are coupled at time $t=0$ (blue line) for some time interval, allowing a heat transfer between them.}
\label{fig:fig1}
\end{figure}

Before we prove Eq.~\eqref{eq:GXFT} let us first discuss a simple example discussed  by Jarzynski and W\'ojcik~\cite{Jarzynski&Wojcik2004}, in fact the simplest case of a GGE system. 
We consider two isolated systems at temperatures $T_1$ and $T_2$, as depicted in Fig.~\ref{fig:fig1}.
 The initial statistics of the combined system  can be written as a GGE, $\rho\propto e^{-\beta_1 E_1-\beta_2 E_2}$, where $E_{1,2}$ are the energy of each subsystem, or equivalently as $\rho\propto e^{-\beta_{+} E_+-\beta_{-} E_{-}}$ with $\beta_{\pm}=\beta_{1}\pm\beta_{2}$, $E_{\pm}=(E_{1}\pm E_{2})/2$.
 Now, at $t=0$ we couple the two systems and examine the fluctuations in the heat exchange $\Delta E_2=-\Delta E_1$ over a given time. Since $E_{+}$ is conserved we get 
$u=\beta_{-}\Delta E_{-}$.
 According to Eq.~\eqref{eq:GXFT}, we must have   $P(\beta_{-}\Delta E_{-})/P(-\beta_{-}\Delta E_{-})=P(\Delta E_{-})/P(-\Delta E_{-})= \exp (\beta_- \Delta E_{-})$. This is exactly the result obtained in Ref.~\cite{Jarzynski&Wojcik2004}. 
 The generalized exchange fluctuation theorem, Eq.~\eqref{eq:GXFT}, that we address in the current letter  may be viewed  as  a situation in which a quasi-integrable system behaves as a set of weakly coupled subsystems, each of which stands for a conserved quantity of the integrable system. The coupling is intrinsic, being the interaction induced by the integrability breaking. 
In the simple example of Fig.~\ref{fig:fig1} the parameter that decides whether fluctuations and flow reversals may or may not be observable is the thinness of the channel, rather than the size of the system: this, in our general setting, will be translated into the magnitude of the breaking of conservation.

%Fig. 3

~\\
{\it Fluctuation Theorem.} ---We now prove the generalized exchange fluctuation theorem given in Eq.~\eqref{eq:GXFT}. \\~\\

{\bf Classical:} \\~\\

  Let us denote the phase-space variables by $\vec{x}=(\vec{q},\vec{p})$, and consider a Hamiltonian $\mathcal{H}=\mathcal{H}_0+\mathcal{H}_p$, where $\mathcal{H}_0$ has conserved quantities $\{J_r\}$, and $\mathcal{H}_p\ll\mathcal{H}_0$ is a perturbation. 
We assume the initial probability distribution is of the GGE form (Eq.~\eqref{GGE}).

 Then, the system evolves from $\vec{x}$ to $\vec{x}^t$ with $\mathcal{H}$ for some time $t$, and we measure the probability of the quantity $u(t)=\sum_r\beta_r (J_r(\vec{x}^t)-J_r(\vec{x}))$.  {\em Note that the $\beta_r$ are the same for initial and final states}. The proof is given under the conditions of (a) time-reversibility of the dynamics, and (b) the functions $J_r(\vec{x})$ are invariant under $(\vec{q},\vec{p})\rightarrow (\vec{q},-\vec{p})$. We use the additional following notations: the operation $(\vec{q},\vec{p})\rightarrow (\vec{q},-\vec{p})$ is designated by $\vec{x}\rightarrow\tilde{\vec{x}}$, and $T(\vec{x}',t;\vec{x})$ denotes the probability of finding the system at state $\vec{x}'$ at time $t$ given that it was at state $\vec{x}$ at time $t=0$. \\~\\

The quantity in question can be written explicitly as
\begin{equation}
P\left(u\right)=\int d\vec{x}d\vec{x}' \delta\left(   \sum_r\beta_r (J_r(\vec{x}')-J_r(\vec{x}))     -u\right)T(\vec{x}',t;\vec{x}) e^{-Q(\vec{x})}/Z.
\label{eq:proof1}
\end{equation}
Exchanging the integration variables $(x \leftrightarrow x')$ and using the delta function to re-express $Q(\vec{x}') \rightarrow Q(\vec{x}) +u$, this becomes:
\begin{equation}
P\left(u\right)=e^{u} \int d\vec{x}d\vec{x}' \delta\left(\sum_r\beta_r (J_r(\vec{x}')-J_r(\vec{x}))+u\right)T(\vec{x},t;\vec{x}') e^{-Q(\vec{x})}/Z.
\label{eq:proof2}
\end{equation}
Time reversibility allows us to preform the transform 
$T(\vec{x}',t;\vec{x})\rightarrow T(\tilde{\vec{x}},t;\tilde{\vec{x}}')$, which, together with the assumed invariance  $J(\vec{x})=J(\tilde{\vec{x}})$ gives the desired result
 $P\left(u(t)=u\right)=e^u P\left(u(t)=-u\right)$. \\

{\bf Quantum:} \\~\\

We  now consider a system where the Hamiltonian almost commutes with a series of operators $\hat J_r$.
Consider the following experiment~\cite{Kurchan2000}:

$\bullet$    Diagonalize the operator $\hat Q = \sum \beta_r \hat J_r$, such that $\hat Q |q \rangle = q  |q \rangle$

$\bullet$  Choose an eigenvector $ |q\rangle$ with probability $e^{-q}$, and evolve it with the complete Hamiltonian for time $t$

$\bullet$ Add the  amplitudes  $\left|\langle q' | e^{-i\frac t \hbar H }  |q \rangle\right|^2$ to the histogram of probabilities of  $u=(q'-q)$, and repeat.

We are in fact calculating
\begin{equation}
P\left(q'-q=u\right)=\int dq  dq' \delta\left(q'-q-u\right)\left| \langle q' | e^{-i\frac t \hbar H }  |q \rangle\right|^2  e^{-q}/Z.
\label{eq:proof5}
\end{equation}
Exchanging as above $(q \leftrightarrow q')$, using the delta function, and the requirement of time-reversal, $H=H^*$  and $|q\rangle =|q \rangle^* $, we easily obtain  the fluctuation relation (\ref{eq:GXFT}).

~\\
{\bf  Coarse-grained constants, the hydrodynamic limit:} \\~\\

The Fluctuation Theorem implies no assumption but as it stands is of little use, for the following reason: it requires a GGE initial condition {\em even for large deviations}, something
that does not follow from any obvious physical process, and if it does, may hardly be expected to be preserved by the dynamics.
Here is where an argument of equivalence of ensembles becomes necessary: we wish to argue that an initial condition may
be considered {\em as if} it were GGE, with some $\beta_r$. 
As is well known in these cases, ensemble equivalence may be expected to  hold for `coarse grained' constants. One further assumes that  these describe correctly the physical situation--- a hydrodynamic limit, the applicability
of which   does not only depend on
the system but also on its initial conditions.  (The issue of equivalence of ensembles in the case of other fluctuation theorems has been discussed previously, e.g., Refs.~\cite{Cleuren_etal2006,Talkner_etal2013,Jeon_etal2015}).
Formally, we may work as follows:  putting $\beta_1=,...,=\beta_{m }\equiv \tilde \beta_1$, then $\beta_{m+1} =,...,=\beta_{2m }=\tilde \beta_2$, etc, we have that 
 $Q= \tilde \beta_1 (J_1+...+J_{m}) + \tilde \beta_2 (J_{m+1}+...+J_{2m})+... $. In other words, considering a situation with $\beta_r$ grouped into sets of $m$ automatically
 yields a coarse-graining on the constants $J_r$, which enter as sums of $m$ terms. The FT clearly works for such an initial condition, as this is only one particular situation.
  For large $m$ we may expect to get equivalence of ensembles, the question that
 remains is whether this grouping we made accurately reflects the original problem, i.e. if a coarse-grained description is faithful. Note however that we are dealing with {\em  large deviations rather than simple averages},
so the kind of equivalence of ensembles we need is very demanding.

Let us consider a more general distribution  $\rho(\vec{x})\propto e^{-Nf(\vec{J}^{cg}(\vec{x}))}$, with $f=O(1)$ and the $\vec{J}^{cg}(\vec{x})$ coarse-grained variables over blocks of size $m=\alpha N$ ($\alpha$ small but $O(1)$) and normalized by $m$ to be of $O(1)$. 
 The dependence of the distribution $f$ on $\vec{J}^{cg}$ reflects the assumption of hydrodynamic limit; it is justified by the fact mentioned above that a classical quasi-integrable system typically visits the approximate torus ergodically in a time much shorter
than the diffusion time.

Then, one may try to see whether  Eq.~\eqref{eq:GXFT} is valid with the specific choice for the values of $\{\beta_r\}$ 
\begin{equation}
\beta_r=\left.\frac{\partial Ns(\vec{J}^{cg})}{\partial J_r}\right|_{\vec{J}_*},
\label{eq:betas}
\end{equation}
where we define the entropy $Ns(\vec{J}^{cg})\equiv\ln\int \delta(\vec{J}^{cg}(\vec{x})-\vec{J}) d\vec{x}$ and $\vec{J}_*$ is a saddle point of $Nf(\vec{J}^{cg})-Ns(\vec{J}^{cg})$, in analogy with standard thermodynamics. We now discuss the limitations of this statement, a full mathematical derivation is given in the Supplemental Material.

Looking at the derivation following  Eqs.~\eqref{eq:proof1}-\eqref{eq:proof2} it is clear that a fluctuation theorem of the form $\ln P(N\Delta f=u)-\ln P(N\Delta f = -u)=u$ can be readily proved. In itself it is not very useful, as we don't have a direct access to $f$. 
The usual FT theorem, for $\beta_r$ defined by (\ref{eq:betas}) would follow if we could identify $N\Delta f\approx\sum\beta_r \Delta J_r^{cg}$. Such an approximation depends both
on the model and on the time-interval considered. 
{ If we assume a large deviation principle for the transition probability for $\vec{J}^{cg}\rightarrow\vec{J}^{cg\prime}$, we  find that  $u$ is dominated by certain values of $J^{cg\prime}_{*}-J^{cg}_{*}$. We have then to assume that these values are bounded to be small, but still very large with respect to their fluctuations of order $1/\sqrt{N}$. Note that this still leaves room for large deviations, because the $\vec{J}^{cg}$ are intensive quantities. See Supplemental Material. }

~\\
{\bf FPU}~\\

{\it An example.} ---We now demonstrate our result by treating a specific quasi-integrable system: the Fermi-Pasta-Ulam-(Tsingou) (FPU) chain~\cite{Dauxois2008}. This is a 1D non-linear chain whose Hamiltonian reads 
\begin{equation}
\mathcal{H}_{\rm FPU}= \frac{1}{2}\sum^N_{n=1} p_n^2 + \sum^N_{n=0} V_{\rm FPU}(q_{n+1}-q_{n}), 
\end{equation}
where $V_{\rm FPU}(r)=r^2/2+A r^3/3+B r^4/4$, and we take a fixed-ends boundary conditions $q_0=q_{N+1}=0$. The dynamics depends on the size of the chain $N$, the energy density $\epsilon=E/N$, and the parameter $B$;  The parameter $A$ can be rescaled by the energy $\epsilon$, and thus is set to $A=1$ hereafter~\cite{Benettin_etal2013}. The FPU potential can be written as a small perturbation of the Toda potential, $V_{\rm Toda}(r)=V_0 (e^{\lambda r}-1-\lambda r)$. For the values $V_0=(2A)^{-2}$ and $\lambda=2 A$ one finds  $\mathcal{H}_{\rm FPU}=\mathcal{H}_{\rm Toda}+\mathcal{H}_{p}$, with $\mathcal{H}_p\sim (2A^2/3-B)\epsilon/4+(A^3/3)\epsilon^{3/2}/5$. The Toda chain is integrable~\cite{Henon1974,Flaschka1974}, and a set of conserved quantities $\{J_r\}$ can be derived, having the properties that:
{\em i)} They are exact constants of the Toda Lattice, {\em (ii)} for weak coupling they are very close to the Fourier modes, except of a small fraction, which may be associated with soliton numbers. See Supplemental Material. 

  The fact that the Toda dynamics serves as an underlying integrable model for the FPU chain has been established in the three main aspects: (1) the presence of solitons in FPU dynamics~\cite{Zabusky&Kruskal1965}, (2) at short timescales, although being chaotic, the FPU dynamics completely explores a Toda tori~\cite{Benettin_etal2013,Benettin_etal2009}, (3) starting form a concentrated ensemble the FPU dynamics drifts between quasi-stable states, each of which can be characterized by Toda tori and a corresponding GGE ensemble~\cite{Goldfriend&Kurchan2018}. 

\begin{figure}
\centerline{\resizebox{0.45\textwidth}{!}{\includegraphics{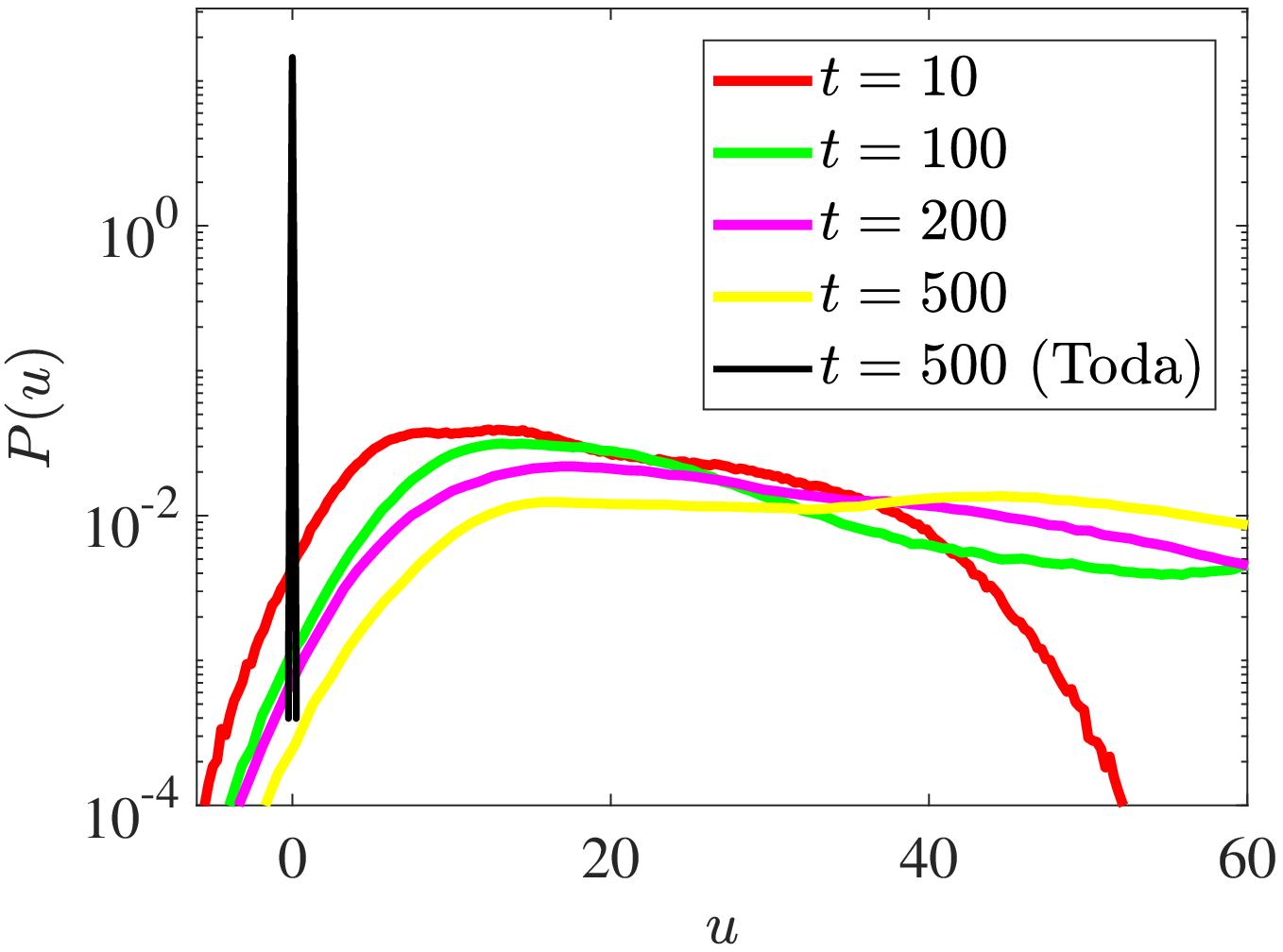}}
%\hspace{2cm}
\resizebox{0.45\textwidth}{!}{\includegraphics{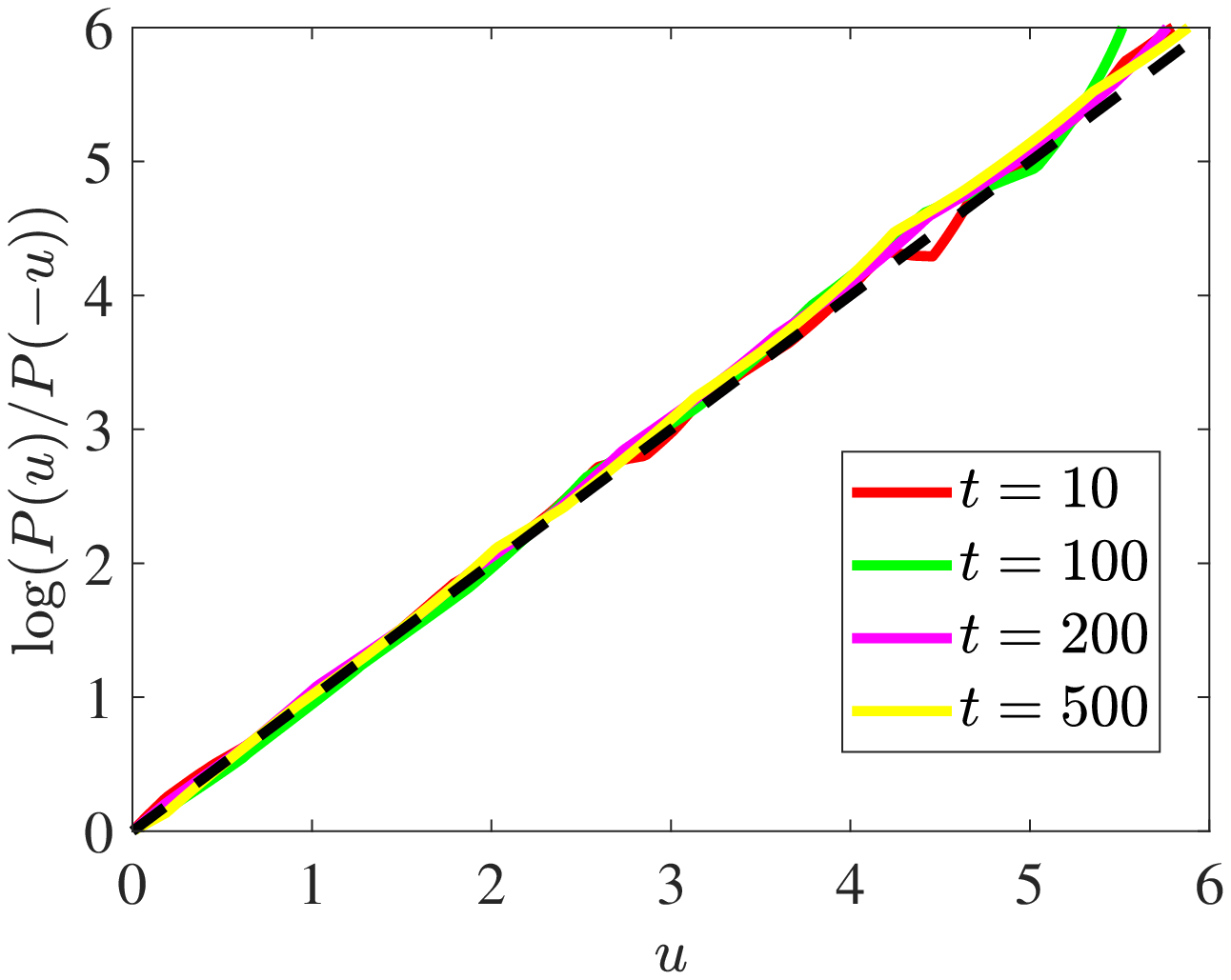}}}
\caption[]{ Illustration of the generalized exchange fluctuation theorem for a FPU chain, with $N=15$. (a) colored: probability distribution of $u=\sum_r \beta_r(J_r(\vec{x}^t)-J_r(\vec{x}))$ at different times when the system evolves with the FPU Hamiltonian. black: the distribution when the system evolves with the Toda Hamiltonian, corresponding solely to random numerical error. (b) verification of the fluctuation theorem (Eq.~\eqref{eq:GXFT}). The time units are set by the frequency of the longest Fourier mode of a linear chain with the same size.}
\label{fig:fig2}
\end{figure}

Fig.~\ref{fig:fig2} demonstrates the GXFT for a system of size $N=15$, with the parameters $A=1$, $B=2/3$. The initial GGE ensemble is generated with a Monte-Carlo sampling. The values of  $\{\beta_r\}$ are chosen to follow a step function profile, $\beta_{1\leq r \leq 6}=\beta_a$ and $\beta_{7\leq r \leq 15}=\beta_b$, with $\beta_b\gg 1$ such that only the lower part of the set is excited and $\langle\epsilon\rangle\approx 0.01$. In Fig.\ref{fig:fig2}a the probability distribution of $u$ is shown for different times. At long times, negative values of the $u$ are rare, indicating a net drift of the conserved quantities as the system approaches toward equipartition. For comparison, we present the distribution of $u$ in the case of the Toda dynamics. Here, any departure from a delta function is only due to a random numerical error. Fig.\ref{fig:fig2}b demonstrates the validity of Eq.~\eqref{eq:GXFT}.

%Fig. 3

~\\
{\bf  Skiing down the free-energy landscape:  evolution of $\beta_r$ } \\~\\

The FT stresses the fact that a coarse-grained system always goes down an (minus)-entropy  landscape, but not necessarily as a gradient, just as
a {\em ski} descent.

Let us distinguish  as above the  conserved $ J^c_1,..., J^c_n$, and non-conserved $J_1,...,J_p$ quantities.
The  inverse temperatures for the two groups are $\beta^c_1,...\beta^c_n$ and $\beta_1,...,\beta_p$, respectively.
Then it is easy to see that in the course of evolution the Lagrange multipliers $\beta^c_1,...\beta^c_n$ will tend to finite values (they impose a constraint for the truly conserved quantities), while  $\beta_1,...,\beta_p$ all go to zero, because  as $t \rightarrow \infty$ they are no longer imposing any constraint. For instance, returning to the simple set-up in Fig.~\ref{fig:fig1}, we know that at long times $T_{1}=T_{2}$, which implies $\beta_{+}\to 2/T_{1}=2/T_{2}$ and $\beta_{-} \to 0$.

Furthermore, the Fluctuation Theorem immediately implies in the usual way that, for $u=\beta_r \Delta J_r$:
\begin{equation}
\langle u \rangle = \int_{-\infty}^{+ \infty}  P(u)\; u \;  du = \int_{0}^{+ \infty} P(u) \; \left[1-e^{-u}\right]\;  u\;  du \geq 0
\end{equation}
This is a form of the Second Principle,  since for a small change $\delta J_r$ we have that 
\begin{equation}
\delta S({\vec{J}}) = \Sigma_r \frac{\partial S}{\partial J_r} \delta J_r = \Sigma_r \beta_r \delta J_r,
\end{equation}
 with $S({\vec{J}})=Ns({\vec{J}})$ being the entropy.

\begin{figure}
\centerline{\resizebox{0.45\textwidth}{!}{\includegraphics{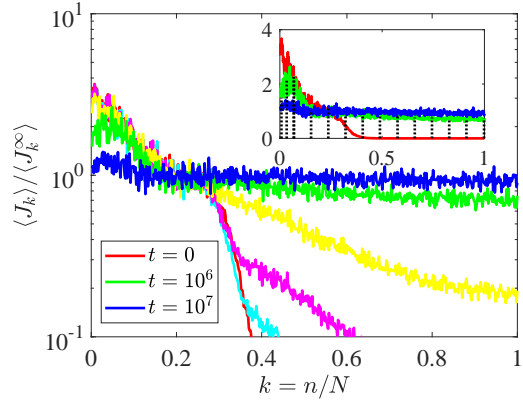}}}
\caption[]{The profile of the Toda constants under FPU dynamics for a system with $N=511$, $\epsilon=10^{-3}$, $A=1$, and $B=2$. The curves are averaged over 102 initial conditions in which only the lowest 0.1 of the Fourier modes are excited. The cyan, magenta, and yellow curves correspond to times $10^3$ ,$10^4$, and $10^5$ respectively. The profiles of $\langle J_k \rangle$ are normalized by the profile obtained in equipartition between all the Fourier modes of a linear chain with the same energy. By the time $2\times 10^7$ the profile fluctuates around 1. Inset: same in linear scale, the dotted black lines represent coarse-graining.}
\label{fig:fig3}
\end{figure}

Finally, we reflect the ideas discussed above in the slow thermalization of a large size FPU chain with low energy. This example will be discussed in detail in a future publication~\cite{Goldfriend&Kurchan2018}. In the modern version of the original FPU numerical experiment one starts with an initial ensemble in which only the lowest Fourier modes are excited and studies its dynamics~\cite{Benettin_etal2013}. Figure~\ref{fig:fig3} shows the time evolution of the profile $\langle J_k\rangle$ for a system with $N=511$ and $\epsilon=10^{-3}$. The time to fill the Toda tori in such system is of order $10^3$, shorter than the typical time for changes in $J_k$ induced by the breaking of integrability. The hydrodynamic limit, which is evident in the self-averaging profiles, suggests that at any time along the dynamics~\footnote{More precisely at times which are longer than the time to fill the initial Toda tori}, the system can be described by a coarse-grained GGE. Indeed, we have verified that this system admits microcanonical ensemble averages with $m=23$ coarse-grained quantities~\cite{Goldfriend&Kurchan2018}.     
Moreover, we see from inset of figure~\ref{fig:fig3} that after some time, the vast majority of the out of equilibrium 
 (linear) variation of quasi-constants of motion is in the lower modes, that we know correspond to the soliton modes (see Supplemental Material). Thus, the evolution of the $\beta_r$ (and $\Delta Q$) at late times describes the gradual death of the excessive solitons, and the Fluctuation Theorem describes the ratio of death and resurrection of those. Here we may note the remark made in the abstract: we could have a macroscopic system with only a few solitons, and the desired fluctuations scale with the number of solitons and not with the size, so they may be observable.\\
 
{\it Acknowledgments.} --- We thank Abhishek Dhar, Leticia Cugliandolo and Jacopo de Nardis for fruitful discussions. TG and JK are supported by the Simons Foundation Grant No. 454943.

\newpage
%% APPENDIX
\section*{Supplemental Material}

{\bf  Other distributions} \\~\\

Let us assume that the system starts from a distribution $\rho$ of coarse-grained variables as above, with $m$ (the coarsening scale) a small but finite fraction of $N$,
and that may  written as $\rho(x) \propto e^{-Nf(\vec{J})}$ ($f$ a smooth function).  For simplicity, we shall drop here the superscript $cg$, but we must
understand coarse graining  implied everywhere. The $\vec{J}$ are now $N/m$   sums of $m$ of the original constants, normalized by $m$. We wish to
discuss under what conditions the FT, which was shown for $f(x)=Q(x)$ may be extended to this more general case, i.e.  ensembles $f$ and canonical are equivalent. 
The quantity we need to analyze is 
\begin{equation}
P(u)=\int d\vec{x}d\vec{x}'\delta\left(\sum\beta_r(J_r(\vec{x}')-J_r(\vec{x}))-u\right) T(\vec{x}',t;\vec{x}) e^{-Nf(\vec{J}(\vec{x}))},
\label{eq:App1}
\end{equation}
where $\beta_r=\left.\frac{\partial Ns(\vec{J})}{\partial J_r}\right|_{\vec{J}_*}$, $Ns(\vec{J})\equiv\ln\int d\vec{x}'' \delta(\vec{J}(\vec{x}'')-\vec{J})$, and $\vec{J}_*$ is a saddle point of $Nf(\vec{J})-Ns(\vec{J})$.

As a first step, we move toward a canonical representation of the above quantity by multiplying Eq.~\eqref{eq:App1} by $\int d\vec{J}d\vec{J}' \delta(\vec{J}(x')-\vec{J}') \delta(\vec{J}(x)-\vec{J})  e^{Ns(\vec{J})}/e^{Ns(\vec{J})}=1$. Preforming now the integral over $\vec{x}$, and $\vec{x}'$, recognizing that the transition probability in terms of the $\vec{J}$  is \\
$T(\vec{J}',t;\vec{J})\equiv
\int d\vec{x}d\vec{x}' T(\vec{x}',t;\vec{x}) \delta(\vec{J}(x')-\vec{J}') \delta(\vec{J}(x)-\vec{J})  /e^{Ns(\vec{J})},
$ 
we find
\begin{equation}
P(u)=\int d\vec{J}d\vec{J}'\delta\left(\sum\beta_r(J'_r-J_r)-u\right) T(\vec{J}',t;\vec{J}) e^{-Nf(\vec{J})+Ns(\vec{J})}.
\label{eq:App2}
\end{equation}
Next, we wish to calculate the above integral with  a saddle point approximation. We assume that $T(\vec{J}',t;\vec{J})=e^{ N tg((\vec{J}'-\vec{J})/t)}$ behaves as a large deviation function. This assumption is plausible, as we consider the large $N$ limit for which at a given time $t$ there is an entropic pressure preferring a specific value of $\vec{J}'-\vec{J}$. After Fourier transforming the delta-function, Eq.~\eqref{eq:App2} can be written as    
\begin{equation}
P(u)=\int d\lambda d\vec{J}d\vec{J}' e^{N\mathcal{I}(\vec{J},\vec{J}',\lambda;\beta_r,u,t)},
\label{eq:App3}
\end{equation}
with
\begin{equation}
\mathcal{I}(\vec{J},\vec{J}',\lambda;\beta_r,u,t)=i\left(\sum\beta_r(J'_r-J_r)-u\right)\lambda/N +t g ((\vec{J}'-\vec{J})/t)-f(\vec{J})+s(\vec{J})=O(1).
\label{eq:AppI}
\end{equation}
The saddle point equations then read
\begin{eqnarray}
\sum\beta_r(J'_r-J_r)&=&u, \label{eq:SPA1}\\
-i\lambda\beta_r/N+\frac{t\partial g ((\vec{J}'-\vec{J})/t)}{\partial J_r}+\frac{\partial s (\vec{J})}{\partial J_r}&=&\frac{\partial f (\vec{J})}{\partial J_r},\label{eq:SPA2}\\
i\lambda\beta_r/N+\frac{t\partial g ((\vec{J}'-\vec{J})/t)}{\partial J'_r}&=&0 \label{eq:SPA3}.
\end{eqnarray}
Thus, the integral in Eq.~\eqref{eq:App2} can be approximated as 
\begin{equation}
\ln P(u)={N \mathcal{I}(\vec{J}_*,\vec{J}'_*,\lambda_*;\beta_r,u,t)}+ O(N^{1/2}),
\label{eq:PuSPA}
\end{equation}
where the subscript $*$ indicate the solution of the saddle point equations. In particular, the solution for $\vec{J}_*$ satisfies  $\left.\frac{\partial f (\vec{J})}{\partial J_{r}}\right|_{\vec{J}_*}=\left.\frac{\partial s (\vec{J})}{\partial J_{r}}\right|_{\vec{J}_*}=\beta_r/N$.

Now, let us calculate the probability in the reverse situation, $P(-u)$. We make use of the expression in Eq.~\eqref{eq:App2} with switching between the dummy variables $\vec{J}$ and $\vec{J}'$, merely for the sake of convenience
\begin{equation}
P(-u)=\int d\vec{J}d\vec{J}'\delta\left(\sum\beta_r(J_r-J'_r)+u\right) T(\vec{J},t;\vec{J}') e^{-Nf(\vec{J}')+Ns(\vec{J}')}.
\label{eq:App2r}
\end{equation}
Since the dynamics has time-reversal invariance, the ratio between the probability of moving from $\vec{J}$ to $\vec{J}'$ and the probability of the reversal dynamics is completely determined by the volume of each initial state, i.e., $T(\vec{J},t;\vec{J}')/T(\vec{J}',t;\vec{J})=e^{Ns(\vec{J}')}/e^{Ns(\vec{J}')}$. Inserting this relation into Eq.~\eqref{eq:App2r} and switching signs within the delta-function we get
$$
P(-u)=\int d\vec{J}d\vec{J}'\delta\left(\sum\beta_r(J'_r-J_r)-u\right) T(\vec{J}',t;\vec{J}) e^{-Nf(\vec{J}')+Ns(\vec{J})},
$$
which can be written as
\begin{equation}
P(-u)=\int d\lambda d\vec{J}d\vec{J}'e^{N\mathcal{I}(\vec{J},\vec{J}',\lambda;\beta_r,u,t)-N(f(\vec{J}')-f(\vec{J}))},
\label{eq:App3r}
\end{equation}
Before preforming a saddle-point approximation for the integral in Eq.~\eqref{eq:App3r}, we write the Taylor expansion
$$f(\vec{J}')-f(\vec{J})=\sum_s\frac{\partial f}{\partial J_s}(J'_s-J_s)+\frac{1}{2}\sum_{s,m}\frac{\partial^2 f}{\partial J_s\partial J_m}(J'_s-J_s)(J'_m-J_m),$$
and assume that $J'_r-J_r=O(a)$ with $a<1$  {\em a small quantity}. The saddle-point equations then read
\begin{eqnarray}
\sum\beta_r(J'_r-J_r)&=&u \label{eq:SPAr1}\\
-i\lambda\beta_r/N+\frac{t\partial g ((\vec{J}'-\vec{J})/t)}{\partial J_r}+\frac{\partial s (\vec{J})}{\partial J_r}&=&O(a^2) \label{eq:SPAr2}\\
i\lambda\beta_r/N+\frac{t\partial g ((\vec{J}'-\vec{J})/t)}{\partial J'_r}&=&\frac{\partial f (\vec{J})}{\partial J_r}+\sum_{s}\frac{\partial^2 f}{\partial J_s\partial J_r}(J'_s-J_s)+O(a^2) \label{eq:SPAr3}.
\end{eqnarray}
Eqs.~\eqref{eq:SPA1}--\eqref{eq:SPA3} are of the form $\nabla \mathcal{I}=0$, whereas Eqs.~\eqref{eq:SPAr1}--\eqref{eq:SPAr3} stand for $\nabla \mathcal{I}=O(a)$. Thus, the solutions to the former and the latter are identical up to order $a$, but give corrections up to order $a^2$ to the evaluation of $\mathcal{I}$.  In addition, the solution yields $f(\vec{J}')-f(\vec{J})=u/N+O(a^2)$.

To sum up,  we get
\begin{equation}
\ln(P(u))-\ln(P(-u))=u+O(a^2),
\end{equation}
hence, the theorem holds for changes in the approximate constants that are fractionally small, but still correspond to a large deviation.\\

{\bf Toda-chain conserved quantities} \\~\\

Below we define a set of quantities, $\{J_r\}$, which are conserved in time under the Toda Hamiltonian, with the potential
\begin{equation}
V_{\rm Toda}(r)=(2A)^{-2}(e^{2A r}-1-2A r).
\end{equation}
The Toda action variables were introduced by Ferguson et. al.~\cite{Ferguson_etal1982}, however, their calculation is intractable. Our choice of the constants of motion is based upon the definition of the exact action variables, capturing their main characteristics. We consider only the case of odd $N$, the case of even $N$ can be readily deduced. Following the analysis of Ref.~\cite{Ferguson_etal1982}, we first give a detailed rigorous procedure for the calculation of the quantities and then indicate their essential properties.

\begin{figure}
\centerline{\resizebox{0.45\textwidth}{!}{\includegraphics{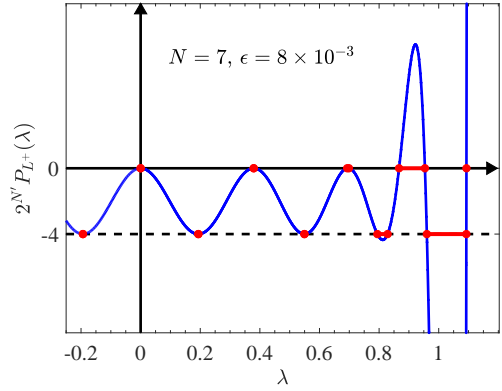}}
}
\caption[]{The characteristic polynomial of the matrix $\vec{L}^+$. The eigenvalues of $\vec{L}^+$ and $\vec{L}^{-}$ are defined by $P_{L^+}(\lambda^+)=0$, and $2^{N'}P_L(\lambda^-)=-4$. The eigenvalues are coming in pairs which are degenerate at zero energy. When the lowest Fourier modes are excited (in the above example only $k=1$ is excited) the degeneracy at the right part of the plot vanishes.} 
\label{fig:figApp1}
\end{figure}

\begin{itemize}
	\item  \textbf{Defining a periodic chain:} We extend the fixed-ends system of $N$ particles into an antisymmetric periodic one with $N'=2(N+1)$ particles:
\begin{equation}
\left\{\begin{array}{l l}
q'_i=q_i,\,\,\,p'_i=p_i, &i=1,\dots,N+1, \\
q'_{N+1+i}=-q_{N+1-i},\,\,\,p'_{N+1+i}=-p_{N+1-i},& i=1,\dots,N+1.	
\end{array}\right.
\label{eq:Per}
\end{equation}
This antisymmetric construction is conserved by the dynamics and implies that the $N'$ conserved quantities of the extended chain are degenerated to $N$ quantities for the fixed-ends one.

\item \textbf{Constructing a Lax matrix:} We define a symmetric matrix $\vec{L}^+$ of size $N'\times N'$,
\begin{equation}
\vec{L}^+=
\begin{pmatrix}
b_1     &a_1     &       &        &        &a_{N'}  \\
a_1     &b_2     &a_2    &        &        &        \\	
        &\ddots  &\ddots &\ddots  &        &        \\	
        &        &\ddots &\ddots  &\ddots  &        \\
        &        &       &a_{N'-2}&b_{N'-1}&a_{N'-1}\\
a_{N'}  &        &       &        &a_{N'-1}&b_{N'}  \\
\end{pmatrix},
$$
where the unoccupied entries are zero and
\begin{equation}
a_n = \frac{1}{2} e^{A(q'_n-q'_{n-1})},\qquad
b_n = A p'_{n-1}.
\end{equation} 
The Hamilton equations for the Toda chain are equivalent to the Lax equation
\begin{equation}
\dot{\vec{L}}^{+}=\vec{B}^{+}\vec{L}^{+}-\vec{L}^{+}\vec{B}^{+}=[\vec{B}^{+},\vec{L}^{+}],
\label{eq:Laxplus}
\end{equation}
with an antisymmetric matrix $\vec{B}^+$, which implies that the eigenvalues of $\vec{L}^+$ do not vary in time under the Toda dynamics~\cite{Flaschka1974,Henon1974}. This underlay the integrability of the Toda Hamiltonian. Note that in particular we have
\begin{equation}
\sum^{N'}_{n=1} (\lambda_n^{+})^{2}=(2A)^2 \mathcal{H}_{\rm Toda}+(N+1),
\label{eq:TodaEnergy}
\end{equation}
where $\lambda_n^+$ are the eigenvalues of $\vec{L}^+$.

\item \textbf{The characteristic polynomial of $\vec{L}^+$:} The action variables of the Toda chain can be defined through the characteristic polynomial of the Lax matrix $\vec{L}^+$, depicted in Fig.~\ref{fig:figApp1}. The polynomial, $P_{L^+}(\lambda)$ , is symmetric about the $y$ axis as a result of the antisymmetric extension to a periodic chain ($\vec{q},\vec{p}\rightarrow\vec{q}',\vec{p}'$). When the chain is at rest, i.e.,  $a_n=1/2$ and $b_n=0$, the extrema of $P_{L^+}(\lambda)$ {\it are degenerate in pairs}, laying on the $x$-axis and the line $y=-4\cdot 2^{-N'}$. The former correspond to the eigenvalues of $\vec{L}^{+}$, the latter are given by the eigenvalues of another matrix $\vec{L}^{-}$~\cite{Kac&Moerbeke1975}
$$
\qquad
L^-_{ij}=
\left\{\begin{array}{l l}
	L^+_{ij}, & (i,j)\neq (1,N), (N,1)\\ 
	-L^+_{ij}, & (i,j)=(1,N), (N,1)\\
\end{array}\right..
\end{equation}
At finite energy, $\epsilon>0$, the degeneracy is broken, as shown in Fig.~\ref{fig:fig2}--- the action variables of the Toda chain are given by integration of a complex function within the corresponding gaps. For simplicity, only the intervals of integration are taken as our choice for the Toda conserved quantities.  

\item \textbf{Definition of $\{J_k\}$:} We know summarize the rigorous procedure for calculating our Toda constants. First, expand the system to a periodic one ($\vec{q},\vec{p}\rightarrow\vec{q}',\vec{p}'$) and construct the matrices $\vec{L}^{+}$ and $\vec{L}^{-}$. Then, calculate and sort the eigenvalues of both matrices in an increasing order. The symmetry of the problem yields the following pattern $-\lambda_1, -\lambda_2,\dots -\lambda_{N'-1},0,0,\lambda_{N'-1},\dots,\lambda_2,\lambda_1$, where we recall that $N'=2N+2$.  All these quantities are conserved by the Toda dynamics since the invariance of the eigenvalues of $\vec{L}^+$ impose the invariance of  $P_{L^+}(\lambda)$. Finally, define  
\begin{equation}
\{J_k\}^{N}_{k=1}=\{\lambda_{2n}-\lambda_{2n+1}\}^{n=N}_{n=1}.
\end{equation} 
\end{itemize}

Let us now emphasize the reasoning of our choice of conserved quantities, indeed other choices have simpler definitions, for instance, the eigenvalues of $\vec{L}^+$ or the trace of $(\vec{L}^+)^p$ with $p=1,\dots,N$. The above definition of $\{J_k\}$ has several important properties~\cite{Ferguson_etal1982}: (1) For small energy, the values of $\{J_k\}$ are correlated with the normal modes of a linear chain--- exciting lower/higher Fourier $k-$modes corresponds to opening gaps (non-degenerated pair of eigenvalues) on the right/left side of the characteristic polynomial. (2) gaps which involve $\lambda_n>1$ correspond to excitation of soliton-like waves. (3) The action variables of the Toda Hamiltonian are given by integrals of the form $\int_{\lambda_{2n+1}}^{\lambda_{2n}} G(P_{L^+}(\lambda)) d\lambda$, where $G$ is a known function. Thus, the number of degrees of freedom, i.e., non-vanishing action variables, correspond to the number of non-vanishing $J_k$.     

\bibliography{QuasiIntegrable_XFT}

\end{document}